\documentclass[preprint,showpacs,aps]{revtex4}
\usepackage{graphicx}
\usepackage{dcolumn}
\usepackage{amsmath}
\usepackage{bm}
\begin{document}
\title {Near-field light emission from nano- and micrometric \\
complex structures}
\author{M. Pieruccini}
\affiliation{CNR, Istituto per i Processi Chimico-Fisici Sez.\ Messina,Via La Farina 237, I-98123 Messina, Italy}
\author{S. Savasta and R. Girlanda}
\affiliation{Istituto Nazionale per la Fisica della Materia (INFM) and Dipartimento di Fisica della Materia e
Tecnologie Fisiche Avanzate, Universit\`{a} di Messina Salita
Sperone 31, I-98166 Messina, Italy}
\author{R. C. Iotti, and F. Rossi}
\affiliation{Istituto Nazionale per la Fisica della Materia (INFM) and Dipartimento
di Fisica, Politecnico di Torino, Corso Duca degli Abruzzi 24, 10129 Torino, Italy}
\date{\today}

\begin{abstract}

We propose a general theoretical scheme for the investigation of light emitted from  nano- and micrometric structures of arbitrary shape and composition. More specifically, the proposed fully three-dimensional approach allows to derive the light-intensity distributions around the emitting structures and their modifications in the presence of nearby scattering objects.
Our analysis allows to better identify the non-trivial relationship between near-field images and fluorescent objects.

\end{abstract}

\maketitle

Recent continuous progress  in scanning near-field microscopy
together with the development of adequate nanofabrication
techniques has enhanced our insight into the field distributions
in the proximity of nano- and microstructured materials. This
insight is of great relevance for the design of future optical
circuitry able to process light with the versatility of electronic
chips. These developments  have stimulated more refined
theoretical approaches as well as novel simulation strategies for
the analysis of light propagation in  complex structures and have
renewed the interest in the classical theory of light
scattering \cite{dereux,madrazo}. Indeed, theoretical predictions
are essential for the design of photonic structures able to
control the propagation of light. As a matter of fact, most of the
theoretical investigations performed so far focus on the
scattering of incident light illuminating structures of
non-trivial shapes \cite{dereux,joanno1}, or study light modes in photonic crystals and
microcavities \cite{joanno}. However, optical circuitry, besides passive
elements controlling/manipulating the flow of photons, require
active components able to emit and/or amplify light. This, in
turn, opens relevant questions concerning the emission patterns of
active mesoscopic systems: Which is the actual field distribution
around emitting structures? How does the interaction with nearby
scattering objects modify the field distribution? What kind of
relationship between near-field images and fluorescent objects
holds?

Aim of this Letter is to provide a general theoretical framework for the evaluation of the field distributions in the proximity of three-dimensional (3D) mesoscopic
fluorescent objects in the presence of nearby scattering structures.
The proposed quantum theory of light emission can be applied to a great variety of
nanostructured optically active materials as
mesoscopic dielectric objects uniformly doped with  optically active molecules (e.g., dye molecules) or
embedding semiconductor  dots or layers able to emit light when appropriately excited.

As a starting point, let us consider the key quantity we want to investigate, i.e., the spectrally-resolved energy density ${\cal I}({\bf r}, \omega)$ corresponding to the electric field at point ${\bf r}$; this can be defined as:
\begin{equation}
\frac{\varepsilon_0}{2} \left< \hat{\bf E}^-({\bf r},\omega) \cdot \hat{\bf E}^+({\bf r},\omega') \right> = {\cal I}({\bf r},\omega) \delta(\omega - \omega')\ ,
\label{e:I}
\end{equation}
where $\hat{\bf E}^+({\bf r},\omega)$ is the electric-field
operator corresponding to the positive frequency $\omega$ (it can be expanded in terms of photon destruction operators) and
$\hat{\bf E}^-({\bf r}, \omega)$ is its Hermitian conjugate.
Our theoretical approach is based on the Green's tensor technique
in the frequency domain \cite{dereux,prasavasta}.
The electric field operator at a given positive
frequency 
can be regarded as the sum of scattering
and emission contributions \cite{prasavasta}:
\begin{equation}
    \hat {\bf E}^+({\bf r}, \omega) = \hat{\bf E}^+_s({\bf r}, \omega) + \hat {\bf E}^+_e({\bf r}, \omega)\ .
\label{se}\end{equation}
The first term describes light coming from free-space and
scattered by the material system, i.e.,
\begin{equation}
     \hat{\bf E}^+_s({\bf r}, \omega) = \hat{\bf E}^+_0 ({\bf r}, \omega) -k^2\int
     \tensor {\bf G}({\bf r}, {\bf r}',{ \omega}) \chi ({\bf r}',\omega)
     \hat {\bf E}^+_0 ({\bf r}', \omega) d{\bf r}'\, ,
    \end{equation}
where $k=\omega/c$, the free-space electric-field operator $\hat {\bf E}^+_0$ describes input light, and
$\chi = \varepsilon - 1$ is the susceptibility function of the material system (that we have assumed  to be a local and scalar function to avoid complications), and $\tensor {\bf G}$ is the Green tensor
obeying the following equation
\begin{equation}
- \nabla \times \nabla \times
\tensor {\bf G}({\bf r}, {\bf r}',{ \omega})
+ k^2 \varepsilon ({\bf r},\omega)
\tensor {\bf G}({\bf r}, {\bf r}',{ \omega})
= {\bf I} \delta({\bf r},{\bf r}')\, ,
\end{equation}
where ${\bf I}$ is the unit dyadic.
We stress that nonlocality and/or anisotropy can be easily implemented \cite{prasavasta}.
The second term in Eq.\ (\ref{se}) describes light emitted by the material system itself, i.e.,
\begin{equation}
 \hat{\bf E}^+_e({\bf r}, \omega) = -i \omega \mu_0 \int
 \tensor {\bf G}({\bf r}, {\bf r}',\omega) \hat{\bf j}({\bf r}', \omega)d{\bf r}'\, ,
\end{equation}
where the integration is performed over the volume of the scattering system and
$\mu_0$ denotes the magnetic permeability of vacuum.
Here, the quantum noise operators
$\hat{\bf j}$ are the sources of spontaneous light emission.
These zero-mean operators can be
derived from the Heisenberg-Langevin equations for the material
system~\cite{henry} and are present only if the susceptibility  tensor
describing the material system
has an imaginary part; they are a direct consequence of the
fluctuation-dissipation theorem and obey the following commutation
rules:\cite{prasavasta}
\begin{equation}
\left[ \hat{j}_l({\bf r},\omega), \hat{j}_{l'}({\bf
r}',\omega)\right] = 0\, , \end {equation}
\begin{equation}
\left[ \hat{j}_l({\bf r},\omega), \hat{j}_{l'}^\dag({\bf
r}',\omega)\right] = \frac{\hbar}{\pi \mu_0}  \frac{\omega^2}{c^2}
|\chi^I({\bf r}, \omega)| \delta_{ll'}\, \delta({\bf r}-{\bf
r}')\, \delta(\omega - \omega')\, , \label{jj}\end {equation}
where $\chi^I$ indicates the imaginary part of $\chi$.
By inserting Eq.~(\ref{se}) into Eq.~(\ref{e:I}), it is easy to
verify that the intensity of the emitted light is directly related
to the correlation function of quantum noise currents:
$\left\langle \hat{\bf j}^\dag \cdot \hat{\bf j} \right\rangle$.
 \begin{figure} \begin{center}\scalebox{1}
{\includegraphics{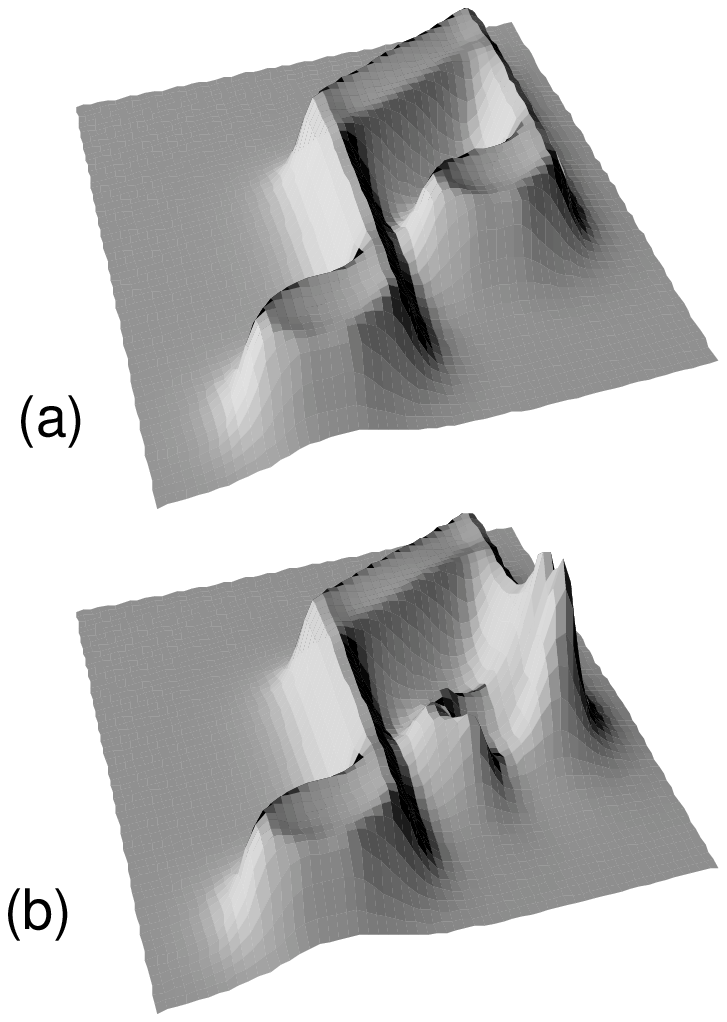}}
 \caption{(a) Field-intensity distribution around a pair of nanoscopic emitting quavers with uniform composition  and population densities; (b) as (a) but with the elliptic pad on the right of the structure replaced with a silver pad. The two calculations have been made at $\lambda = 600$ nm and on the observation plane located 30 nm
above the structure.}
 \label{fig1}\end{center}
 \end{figure}
In order to calculate it explicitly we adopt  a simple 2-level molecular model to describe the  dielectric function of the scattering system,
\begin{equation}
\chi ({\bf r},\omega) = \chi_s({\bf r}) + {\cal N}({\bf r})
\frac{d^2}{\varepsilon_0 \hbar}
\frac{N_a({\bf r}) -N_b({\bf r}) }{\omega_0-\omega +i \gamma}\, ,
\end {equation}
where $\chi_s$ is the real susceptibility of the scattering system
in the absence of active molecules,
 $d$ is the transition dipole moment, ${\cal N}$ gives the density of active molecules, and $N_{a(b)}$ indicates the population densities of the upper (a) and lower (b) levels. This simple model can be easily extended to include non-random molecular orientations,  more energy levels and
molecular species. Also more complex models describing, e.g.,
semiconductor dielectric functions, \cite{henry} can be adopted.
By using this simple model we obtain,
\begin{equation}
    \left\langle \hat{j}^\dag_l({\bf r},\omega) \hat{j}_{l'}({\bf
    r}',\omega')\right> =
    \frac{\hbar}{\pi \mu_0}  \frac{\omega^2}{c^2} \chi^I ({\bf r},\omega)
N({\bf r})
    \delta_{ll'}\, \delta({\bf r}-{\bf r}')\,
    \delta(\omega - \omega')\, ,
\label{jdagj2}
\end{equation}
with the population factor
$N({\bf r}) = N_a({\bf r}) / (N_b({\bf r})-N_a({\bf r}))$. 
In the following we consider the case of absence of input photons
($\left\langle {\bf E}_0^-({\bf r},\omega) \cdot {\bf E}_0^+({\bf
r},\omega') \right\rangle = 0$). By combining Eqs.~(\ref{se}) and
(\ref{jdagj2}), we obtain
${\cal I}({\bf r}, \omega) = \sum_l  {\cal I}_l({\bf r}, \omega)$ with
\begin{equation}
{\cal I}_l({\bf r}, \omega) = \frac{\hbar k^4}{2 \pi}  \sum_{l'}
\int N({\bf r}')\, \varepsilon^I ({\bf r}',\omega)
\left|G_{ll'}({\bf r}, {\bf r}', \omega) \right|^2 d {\bf r}' \ .
\label{emission}\end{equation} We observe that population
densities in principle are affected by light emission, thus Eq.\
(\ref{emission}) should be solved self-consistently together with
the rate equations for the populations. However, in  many cases
(far from the laser threshold), population densities are poorly
affected by light emission and propagation, being fixed by
external pumping. We also observe that Eq.\ (\ref{emission}) has a structure analogous
to the corresponding equation describing emission from thermal sources \cite{greffet,greffet1}. 
 \begin{figure} \begin{center}\scalebox{1}
{\includegraphics{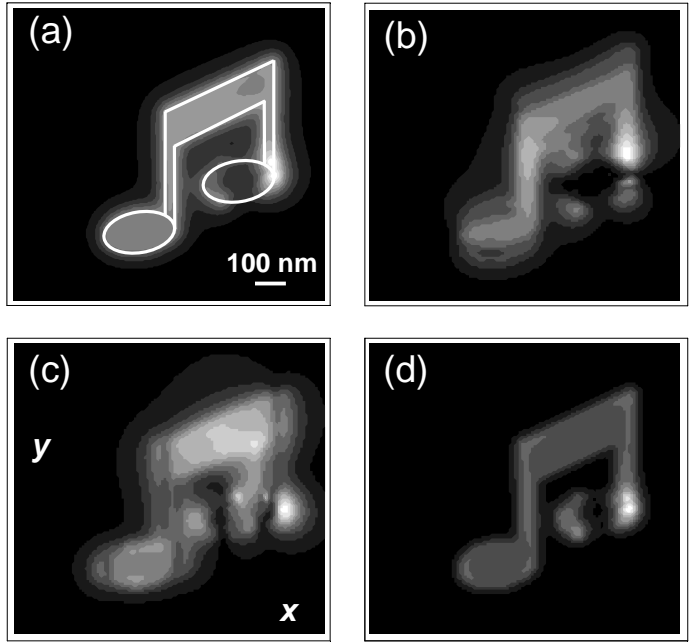}}
 \caption{(a) Field-intensity distribution around a pair of nanoscopic emitting quavers with uniform composition  and population densities; (b) as (a) but with the elliptic pad on the right of the structure replaced with a silver pad. The two calculations have been made at $\lambda = 600$ nm and on the observation plane located 30 nm
above the structure.}
 \label{fig2}\end{center}
 \end{figure}
 \begin{figure} \begin{center}\scalebox{1}
{\includegraphics{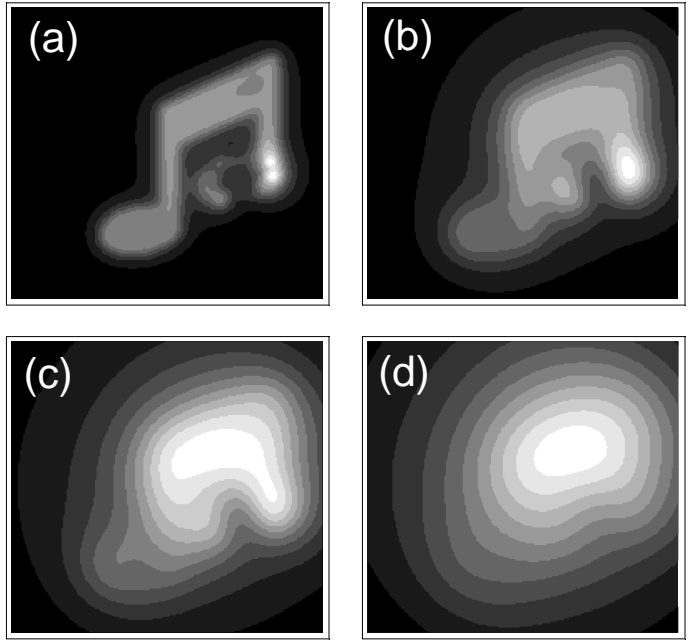}}
 \caption{The field intensity ${\cal I}({\bf r}, \omega)$ (a) and partial intensities ${\cal I}_l({\bf r}, \omega)$  for the structure with the metallic pad; (b) ${\cal I}_x$, (c) ${\cal I}_y$, (d) ${\cal I}_z$.}
 \label{fig3}\end{center}
 \end{figure}

To assess the power and versability of the proposed scheme, we
investigate the field-intensity distribution around a fluorescent low-simmetry nano-structured system.
We consider first a pair of 3D emitting quavers with uniform composition (relative dielectric constant $\varepsilon =8.9 + 0.5i$) and population densities. 
The size of the whole structure is $450 \times 380$ nm and its height is  30 nm. 
Fig.\ 1 (a) displays ${\cal I}(x,y,\bar z, \omega)$  in the proximity of the emitting object in an observation plane located at $\bar z =$ 30 nm. The Green tensor is computed by solving numerically a discretization of the Dyson equation
\cite{dereux}. Calculations have been made at $\lambda = 2 \pi c/\omega = 600$ nm. The emission pattern reproduces almost perfectly the object shape (see also Fig.\ 2a where the structure is outlined). The light intensity is highest at the structure edges and nearby the smaller structures.
Fig.\ 1b shows the effect of replacing the elliptic pad on the right of the structure with a silver 
pad. The emission pattern results to be significantly altered. In particular a strong enhancement of the field intensity nearby the edge of the silver pad is clearly observed while intensity is strongly depleted towards the pad center. This behaviour is typical of surface plasmons \cite{plasmon}, in this case excited by the incoherent light originating from the nearby fluorescent structure.

The vector character  of the emitted light is shown in Fig.\ 2. It displays the partial intensities ${\cal I}_l({\bf r}, \omega)$ (i.e. the intensities that can be detected by a probe able to select light polarization) for the structure with the metallic pad. In spite of the isotropic character of the source currents, the three intensity distributions strongly differ one from each other. We observe that only  ${\cal I}_z$ reproduces quite well the object shape (except the metallic pad of course). Fig.\ 3 displays  ${\cal I}(x,y,\bar z, \omega)$ for (a) $\bar z=30$, (b) $\bar z=60$, (c) $\bar z=90$, and (d) $\bar z=150$ nm above the object. It shows how the  relationship between near-field images and fluorescent objects is progressively lost at increasing observation distances.

We have proposed a general theoretical scheme for the investigation of light emitted from  nano- and micrometric structures of arbitrary shape and composition. The presented numerical results provide useful guidelines for the interpretation of near-field photoluminescence spectroscopy/microscopy measurements. Moreover this theoretical and numerical scheme can be applied to the design of photonic systems with active elements.


We thank Omar Di Stefano for stimulating discussions.


\newpage
\noindent {\bf FIGURE CAPTIONS}
\newline
\newline
\newline
\noindent {\bf Fig.\ 1}
\newline
(a) Field-intensity distribution around a pair of nanoscopic emitting quavers with uniform composition  and population densities; (b) as (a) but with the elliptic pad on the right of the structure replaced with a silver pad. The two calculations have been made at $\lambda = 600$ nm and on the observation plane located 30 nm
above the structure. \newline
\newline
 \noindent {\bf Fig.\ 2}
\newline
The field intensity ${\cal I}({\bf r}, \omega)$ (a) and partial intensities ${\cal I}_l({\bf r}, \omega)$  for the structure with the metallic pad; (b) ${\cal I}_x$, (c) ${\cal I}_y$, (d) ${\cal I}_z$.
\newline
\newline
 \noindent {\bf Fig.\ 3}
\newline
The field intensity ${\cal I}$ calculated on observation planes at increasing distance from the structure; (a) z= 30 nm, (b) z= 60 nm, (c) z= 90 nm, (d) z= 150 nm.

\end{document}